\documentclass[aps,prd,twocolumn,superscriptaddress,showpacs,amsmath,amssymb,amsthm,nofootinbib, preprintnumbers]{revtex4-2}
 \usepackage{amsmath}
\usepackage{graphicx}
\usepackage{epstopdf}
\usepackage{float}
\usepackage{hyperref}
\usepackage{color}
\usepackage[T1]{fontenc}
\usepackage[utf8]{inputenc}
\usepackage[toc,page]{appendix}
\usepackage[usenames,dvipsnames]{xcolor}
\usepackage[normalem]{ulem}
\usepackage{lipsum, babel}
\usepackage[justification=justified,format=plain]{caption}

\usepackage{enumitem}   
\usepackage{subcaption}
\usepackage[utf8]{inputenc}
\usepackage{url}
\usepackage{xcolor}
\usepackage{physics}

\usepackage{hyperref}

\usepackage{graphicx}
\usepackage{dcolumn}
\usepackage{bm}

\newcommand{\be}{\begin{equation}}             
\newcommand{\ee}{\end{equation}}               
\newcommand{\ba}{\begin{eqnarray}}
\newcommand{\ea}{\end{eqnarray}}

\begin{document}

\title{Cardy Entropy of Charged and Rotating Asymptotically AdS and Lifshitz Solutions with a Generalized Chern--Simons term}
\author{Moises Bravo-Gaete}
\email{mbravo@ucm.cl}
\affiliation{Departamento de Matem\'atica, F\'isica y Estad\'istica, Facultad de Ciencias
B\'asicas, Universidad Cat\'olica del Maule, Casilla 617, Talca, Chile.}

\author{Adolfo Cisterna}
\email{adolfo.cisterna@mff.cuni.cz}
\affiliation{Sede Esmeralda, Universidad de Tarapac{\'a}, Avenida Luis Emilio Recabarren 2477, Iquique, Chile}
\affiliation{Institute of Theoretical Physics, Faculty of Mathematics and Physics,
Charles University, V Holešovickách 2, 180 00 Prague 8, Czech Republic}

\author{Mokhtar Hassaine}
\email{mokhtar.hassaine@gmail.com}
\affiliation{Instituto de Matem\'atica, Universidad de Talca, Casilla 747, Talca, Chile}

\author{David Kubiz\v n\'ak}
\email{david.kubiznak@matfyz.cuni.cz}
\affiliation{Institute of Theoretical Physics, Faculty of Mathematics and Physics,
Charles University, V Holešovickách 2, 180 00 Prague 8, Czech Republic}
\date{June 5, 2025}

\begin{abstract}
We consider a three-dimensional gravity model that includes (non-linear) Maxwell and 
Chern--Simons-like terms, allowing for the existence of electrically charged rotating black hole solutions with a static electromagnetic potential. We verify that a Cardy-like formula, based not on central charges but on the mass of the uncharged and non-spinning soliton, obtained via a double Wick rotation of the neutral static black hole solution, accurately reproduces the Bekenstein--Hawking entropy.
Furthermore, we show that a slight generalization of this model, incorporating a dilatonic field and extra gauge fields, admits charged and rotating black hole solutions with asymptotic Lifshitz behavior. The entropy of these solutions can likewise be derived using the Cardy-like formula, with the Lifshitz-type soliton serving as the ground state. 
Based on these results, we propose a generalized Cardy-like formula that successfully reproduces the semiclassical entropy in all the studied cases.
\end{abstract}

\maketitle

\section{Introduction}
Black holes have long stood as pivotal objects in theoretical physics, not only for their gravitational peculiarities but foremost for their unexpected thermodynamic behavior.
The fact that they possess a temperature and an entropy implies, almost inevitably, the need for a statistical interpretation of these thermodynamic quantities -- an interpretation grounded in the counting of underlying microscopic states. This perspective becomes all the more natural within the framework of quantum gravity, where the emergence of thermodynamic laws hints at deeper microscopic dynamics yet to be fully understood.

The formulation of the Anti-de
Sitter/Conformal Field Theory (AdS/CFT) correspondence  \cite{Maldacena:1997re} offered a powerful lens through which to approach this issue. If gravity in an asymptotically AdS spacetime can be described in terms of a CFT on the boundary, then it becomes meaningful to seek a microscopic derivation of black hole entropy directly from the dual CFT.

A key foundational step in this direction was provided by the seminal work of Brown and Henneaux \cite{Brown:1986nw}, who established that the asymptotic symmetry group of a three-dimensional AdS space enlarges to two copies of the Virasoro algebra with a well-defined central charge. This discovery not only revealed a hidden conformal structure at the boundary but also signaled a possibility for a two-dimensional conformal field theory to govern the 
low-energy dynamics of a gravitational system. 
This result encapsulated the core ideas of the AdS/CFT correspondence and opened a concrete path towards the statistical understanding of black hole entropy, particularly in three-dimensional gravity, where black holes such as the Bañados--Teitelboim--Zanelli (BTZ) solution \cite{Banados:1992wn} can be successfully described by employing 
CFT methods.

A crucial development followed from Strominger \cite{Strominger:1997eq}, who applied the well-known Cardy formula \cite{Cardy:1986ie} to the BTZ black hole \cite{Banados:1992wn}, successfully reproducing its semiclassical entropy. This provided a concrete example where the black hole entropy could be calculated microscopically through CFT techniques. Similar computations have been extended to higher-dimensional black holes in general relativity that possess a two-dimensional CFT dual, as well as to strongly coupled field theories with an AdS dual. In the latter case, the Cardy--Verlinde formula relates the entropy of a CFT with a large central charge to its total energy and Casimir energy, offering further insights into the thermodynamic properties of black holes \cite{Verlinde:2000wg}.

However, an extension of this approach to three-dimensional gravity with additional matter fields, particularly in the so-called ``hairy sector'', requires careful considerations. It has been shown that a proper choice of the ground state, identified as a gravitational scalar soliton, plays a fundamental role in correctly reproducing the black hole entropy \cite{Correa:2010hf, Gonzalez:2011nz}. By reformulating the Cardy formula in terms of vacuum energy rather than central charges and identifying the vacuum energy with the mass of the soliton, this method successfully recovers the semiclassical entropy. Notably, this soliton-based approach does not rely on asymptotic symmetries or central charges, making it applicable to a broader class of black hole solutions, including anisotropic spacetimes.

Further developments have extended the soliton method to higher dimensions and various black hole geometries, particularly those with flat horizons; see, e.g., \cite{BravoGaete:2017dso, Alkac:2024hvu}. In such cases, the soliton can be systematically derived from the black hole solution via a double Wick rotation and by appropriate rescaling. This ensures a smooth, regular ground state configuration, reinforcing its role in the microscopic description of black hole entropy.

Extensions of Cardy-like formulas to black holes that are not asymptotically AdS have also been explored, particularly in the case of anisotropic Lifshitz geometries \cite{Gonzalez:2011nz}, with various applications discussed in the literature; see, e.g. \cite{Ayon-Beato:2015jga, Ayon-Beato:2019kmz}. Similar investigations have also been carried out for hyperscaling-violating black holes \cite{Bravo-Gaete:2015wua}, further broadening the scope of these entropy formulas beyond the AdS framework.

The aim of this paper is twofold.  First, we construct a wide variety of rotating and charged black hole backgrounds with abundant matter fields and with various asymptotic symmetries, and study their thermodynamics. Second, based on analyzing the Smarr relations for these solutions, we propose a generalized Cardy-like formula, see Eq. \eqref{entropy} below, that correctly reproduces their entropy, employing the mass of the corresponding soliton ground state. 

More concretely, in the next section, we start from three-dimensional models of Einstein gravity with a negative cosmological constant, coupled to both linear and nonlinear \cite{Hassaine:2007py, Hassaine:2008pw} electrodynamics, and augmented by a Chern--Simons-like term, with the latter allowing for finding rotating solutions characterized by an electrostatic vector potential \cite{Deshpande:2024vbn,Hale:2024zvu}. Within this setup, we construct a variety of charged and rotating black hole solutions with asymptotically AdS behavior and carefully study their thermodynamics, focusing in particular on deriving various Smarr relations. In Sec.~\ref{sec:3}, by extending the previous model to include a dilaton field and by doubling the Chern--Simons-like term, we succeed in generating novel rotating black hole solutions with asymptotically Lifshitz symmetries and study their thermodynamics.  In both instances, the corresponding soliton solution is obtained via a double Wick rotation of the uncharged and non-rotating configuration, and its mass is identified. The obtained thermodynamic relations are then used in Sec.~\ref{sec:4} to propose a generalized Cardy-like formula \eqref{entropy} which correctly reproduces the entropy of all the black hole configurations considered in this paper.
Sec.~\ref{sec:5} is devoted to conclusions. Appendix~\ref{quasilocal} provides further details on the computation of soliton masses using the quasi-local method.

\section{Charged and rotating AdS solutions and their Cardy entropy}\label{sec:2}

\subsection{Action with Chern--Simons-like terms}

We consider the standard three-dimensional Einstein–Hilbert action with a negative cosmological constant, coupled to a (non-)linear Maxwell term \cite{Hassaine:2007py, Hassaine:2008pw} to which we incorporate a Chern--Simons-like term \cite{Deshpande:2024vbn, Hale:2024zvu}. The action reads:
\begin{align}
 S =& \frac{1}{16\pi}  \int d^3x\sqrt{-g}\left(R+\frac{2}{\ell^2}-\beta_p\left(F_{\mu\nu}F^{\mu\nu}\right)^p \right) \nonumber \\
 &+ \frac{\lambda}{4\pi}\int A\wedge H\wedge K\,,
 \label{action1}
\end{align}
where $F_{\mu\nu}=\partial_{\mu}A_{\nu}-\partial_{\nu}A_{\mu}$ is the field strength of the electromagnetic potential $A_{\mu}$, $p$ is a parameter that codifies the nonlinear character of the Maxwell electrodynamics, and $\beta_p$ is the corresponding dimensionful coupling with dimensions $[\beta_p]=L^{2(p-1)}$. Of course, the case $p=1$ ($\beta_1=1)$ will correspond to the standard linear Maxwell electrodynamics, whereas $p=\frac{3}{4}$ describes the conformal case \cite{Hassaine:2007py} \footnote{We
are aware that working with fractional powers $p$ may lead to complex-valued solutions due to the presence of the term $(F_{\mu\nu}F^{\mu\nu})^p$ in the Lagrangian. However, one can always consider the option of introducing a minus sign into the parenthesis, or alternatively, to take the absolute value, in order to ensure the reality of the resultant expressions, e.g. \cite{Cardenas:2014kaa}.}. 
In addition, 
$H$ and $K$ are two one-form fields defined as  $H=dB$ and $K=dC$, with  $B$ and $C$ two non-dynamical $0$-form fields, and $\lambda$ is the coupling constant accompanying the generalized Chern--Simons term; here we take the fields $A_\mu, B, C$ as well as the coupling $\lambda$ to be all dimensionless. By varying the metric, we obtain the Einstein field equations
\begin{equation}\label{EE}
G_{\mu\nu}-\frac{1}{\ell^2} g_{\mu\nu}=8\pi T_{\mu\nu}\,,
\end{equation}
where the energy–momentum tensor associated with the nonlinear Maxwell field is given by
\begin{equation}\label{Tmunu-nonlinear}
T_{\mu\nu}=\frac{\beta_p}{4\pi}\Bigl(p F_{\mu\alpha}F_{\nu}{}^\alpha \left(F_{\mu\nu}F^{\mu\nu}\right)^{p-1}-\frac{1}{4}g_{\mu\nu} \left(F_{\alpha\beta}F^{\alpha\beta}\Bigr)^p\right)\,.
\end{equation}
The corresponding modified Maxwell equations take the following form:
\begin{equation}\label{FJ-nonlinear}
\nabla_\mu \Big(p\beta_p  \left(F_{\alpha\beta}F^{\alpha\beta}\right)^{p-1}F^{\mu\nu}\Big)=\lambda *(H\wedge K)^{\nu}=J^{\nu}\,,
\end{equation}
where the current $J^{\nu}$ is constructed from the field strengths $H$ and $K$, satisfying the algebraic constraints
\begin{equation}\label{algebraic}
F\wedge H=0\,, \quad F\wedge K=0\,.
\end{equation}
We will look for solutions with arbitrary $p$ within the following metric ansatz: 
\begin{equation}\label{BTZ1}
ds^2=-N^2(r)f(r) dt^2+\frac{dr^2}{f(r)}+r^2(d\varphi+h(r) dt)^2\,.
\end{equation}

\subsection{Maxwell electrodynamics \label{linearcase}}
As shown in  \cite{Hale:2024zvu},
in the standard Maxwell case $p=1$ and $\beta_1=1$, the inclusion of the generalized Chern--Simons term allows for  the original charged BTZ solution \cite{Banados:1992wn} to satisfy the field equations. The metric solution within the ansatz \eqref{BTZ1} reads 
\begin{align}\label{BTZ0}
f(r) &= \frac{r^2}{\ell^2}+\frac{j^2}{r^2}-m-2q^2\log\left(\frac{r}{r_q}\right)\,, \nonumber \\
h(r) &= \frac{j}{r^2}\,, \quad N(r)=1\,,
\end{align}
while the gauge field and auxiliary fields take the form
\begin{equation}\label{BTZ2}
A= q\log\left(\frac{r}{r_q}\right) dt\,, \quad B(t,r)=\frac{t}{\lambda r}\,, \quad C(r)=-\frac{2jq}{r}\,.
\end{equation}
Here, $m$ and $q$ are dimensionless integration constants related to mass and electric charge, $j$ has dimensions of length and describes the angular momentum, and $r_q$ is an ``arbitrary scale''. 
Despite the presence of rotation, note the presence of an electrostatic vector potential $A$, which can only exist due to the non-trivial current induced by the auxiliary fields $B$ and $C$.

The thermodynamic quantities associated with this solution can be easily computed. In details, the mass ${\cal M}$, entropy ${\cal S}$, angular momentum ${\cal J}$ and the electric charge ${\cal  Q}_e$ are given by 
\begin{eqnarray}
{\cal M}&=& {\frac{m}{8}}=\frac{r_h^2}{8\ell^2}+\frac{j^2}{8 r_h^2}-\frac{q^2}{4}\log\left(\frac{r_h}{r_q}\right)\,,\nonumber \\
{\cal J}&=&\frac{j}{4}\,, \quad {\cal Q}_e=\frac{q}{2}\,, \quad {\cal S} = \frac{\pi r_h}{2},
\label{charges} 
\end{eqnarray}
where $r_h$ is the horizon location, which corresponds to the largest root of the metric function $f$. The temperature $T$ and the chemical potentials, namely the angular velocity $\Omega$ and the electrostatic potential $\phi_e$ read
\begin{align}
 T&={\frac{f'(r_h)}{4\pi}}=\frac{r_h^4-j^2\ell^2-q^2\ell^2r_h^2}{2\pi r_h^3 \ell^2}\,, \\
 \Omega&=\frac{j}{r_h^2}\,,  \quad \phi_e=-q\log\left(\frac{r_h}{r_q}\right)\,.
\end{align}
These satisfy the standard first law of thermodynamics
\begin{equation}\label{first-law}
\delta {\cal M}=T\delta  {\cal S}+\phi_e \delta  {\cal Q}_e+\Omega \delta  {\cal J}\,.
\end{equation}
However, to satisfy the Smarr relation obtained from Euler scaling of thermodynamic quantities, one would need to consider the pressure/volume or $P/V$ term, as well as a conjugate quantity to $r_q$, namely $\Pi_{r_q}$. These read 
\be
P=\frac{1}{8\pi \ell^2}\,,\quad V=\pi r_h^2\,,\quad \Pi_{r_q}=\frac{q^2}{4r_q}\,, \label{eq:pres-vol}
\ee 
and account for ``extra parameters'' of the solution $\ell$ and $r_q$. In their presence, the above first law generalizes to 
\be 
\delta {\cal M}=T\delta  {\cal S}+\phi_e \delta  {\cal Q}_e+\Omega \delta  {\cal J}+V\delta P+\Pi_{r_q}\delta r_q\,,\label{eq:first-law-ext}
\ee 
with the standard first law above recovered by fixing $\ell$ and $r_q$. It is now easy to check that upon the following (dimensional) scalings:
\be 
r_h\to \alpha r_h\,,\  \ell\to \alpha \ell\,,\  j\to \alpha j\,,\  q\to \alpha^0 q\,,\  r_q\to \alpha  r_q\,,
\ee 
the above thermodynamic quantities scale as
\begin{eqnarray} 
&&{\cal M}\to \alpha^0 {\cal M}\,,\ {\cal J}\to \alpha {\cal J}\,,\ {\cal Q}_e\to \alpha^0 {\cal Q}_e\,,\nonumber\\
&&\ {\cal S}\to \alpha {\cal S},\,P\to \alpha^{-2}P\,.
\end{eqnarray}
Via the Euler scaling argument for homogeneous functions, this then results in the following Smarr relation:
\be 
2PV=T{\cal S}+\Omega{\cal J}+\Pi_{r_q} r_q.\label{eq:PV}
\ee 
Alternatively, {by not scaling} the horizon radius $r_h$, that is 
\begin{eqnarray}\label{eq:scaling}
&& r_h\to \alpha^0 r_h\,,\  \ell\to \alpha^{-1} \ell\,,\  j\to \alpha j, \nonumber\\  
&& q\to \alpha q\,,\  r_q\to \alpha^0  r_q\,,
\end{eqnarray}
we have 
\begin{eqnarray} 
&&{\cal M}\to \alpha^2 {\cal M}\,,\ {\cal J}\to \alpha {\cal J}\,,\ {\cal Q}_e\to \alpha {\cal Q}_e, \nonumber\\ &&{\cal S}\to \alpha^0 {\cal S}\,,\ P\to \alpha^{2}P\,,
\end{eqnarray}
which yields the following Smarr relation: 
\be 
{\cal M}=\frac{1}{2}\left(\Omega{\cal J}+\phi_e {\cal Q}_e+2PV\right)\,.
\label{smarr1}
\ee 
Finally, combining eqs. (\ref{eq:PV}) and (\ref{smarr1}), we recover 
\be 
{\cal M}=\Omega {\cal J}+\frac{1}{2}(T{\cal S}+\phi_e {\cal Q}_e+\Pi_{r_q} r_q)\,.
\ee 
Note the presence of the last term induced by the logarithmic behavior of the
Maxwell field in (2+1) dimensions.

\subsection{Non-linear   electrodynamics \label{nonlinearcase}}
In the pure gravity theory with a non-linear Maxwell electrodynamics, charged black hole solutions were derived in \cite{Hassaine:2007py,Hassaine:2008pw}, where their thermodynamic properties were analyzed in \cite{Gonzalez:2009nn}.
Here, we consider the three-dimensional case with rotation and with the Chern--Simons-like term. In this case, for arbitrary $p$, a class of rotating solutions consistent with the ansatz  \eqref{BTZ1} is given by 
\begin{align}\label{nonlinear-metric}
f(r) &= \frac{r^2}{\ell^2}+\frac{j^2}{r^2}-m+\frac{ (-8)^{p} q^{2p}(p-1)^{2p-1}}{2\, \beta_p^{\frac{1}{2p-1}} (2p-1)^{2(p-1) }r^{\frac{2(1-p)}{2p-1}}}\,, \nonumber \\
h(r) &= \frac{j}{r^2}\,, \quad N(r)=1\,.
\end{align}
Here, the gauge and auxiliary fields supporting the above configuration take the following form:
\begin{eqnarray}\label{BTZ2}
A_\mu dx^\mu&=& \frac{q}{\beta_p^{\frac{1}{2p-1}}}\left(r^{\frac{2(p-1)}{(2p-1)}}-r_h^{\frac{2(p-1)}{(2p-1)}}\right)dt\,, \quad B(t,r)={\frac{t}{\lambda r}}\,, \nonumber\\
C(r)&=&{\frac{p (-8)^p j}{2r} \left(\frac{q (p-1)}{2p-1}\right)^{2p-1}}\,.
\end{eqnarray}
As before, $m$ and $q$ are dimensionless integration constants related to mass and charge, and $j$ has dimensions of length and describes the angular momentum of the hole. 
To ensure AdS asymptotics,  $p$ must obey $2(p-1)/(2p-1)<2$, that is, one has to have  
$p\in (-\infty, 0\,] \cup (\frac{1}{2}, +\infty)$.
It is interesting to note that the limit $p=1$ reduces to the uncharged BTZ solution with trivial fields, even though a solution with $p=1$ and non-trivial fields does exist, as found in \cite{Hale:2024zvu} and reviewed in Sec. \eqref{linearcase}. This means that the previous solution with $p\not=1$ is disconnected from the standard $p=1$ charged case.
Note also that the conformal case, $p=\frac{3}{4}$, was previously studied in details in \cite{Hale:2024zvu} \footnote{Here, as before and as shown in \cite{Cardenas:2014kaa}, to ensure the reality of the expressions, we can introduce a minus sign into the parenthesis or take the absolute value of the term $(F_{\mu\nu}F^{\mu\nu})^p$ from the action (\ref{action1}).}.

We now turn to the thermodynamic properties of this solution. The extensive thermodynamic quantities, derived from the geometric and gauge parameters, are given by
\begin{eqnarray}
\label{Nonqtes}
\mathcal{M}&=&\frac{m}{8}=\frac{r_{h}^{2}}{8\ell^{2}}+\frac{j^2}{8 r_{h}^{2}}+\frac{(-8)^{p} q^{2p}(p-1)^{2p-1}}{16 \beta_p^{\frac{1}{2p-1}} (2p-1)^{2(p-1)} r_h^{\frac{2(1-p)}{2p-1}}}\,,\nonumber\\
\mathcal{S}&=&\frac{\pi r_h}{2}\,,\quad 
\mathcal{J}=\frac{j}{4}\,,\nonumber\\
\mathcal{Q}_e&=& {\frac{p (-8)^p}{8}  \left( \frac{q(p-1)}{2p-1} \right)^{2p-1}.}
\end{eqnarray}
The corresponding intensive (conjugate) thermodynamic parameters are:
\begin{eqnarray}
T&=&{\frac{f'(r_h)}{4\pi}}\nonumber\\
&=&\frac{r_{h}}{2 \pi \ell^{2}}-\frac{j^2}{2 \pi r_{h}^{3} }
+{\frac{(-8)^p \, q^{2p} \, (p - 1)^{2p} }{4 \pi \beta_p^{\frac{1}{2p-1}}(2p - 1)^{2p-1} \, r_h^{\frac{1}{2p - 1}}}} \,,\nonumber\\
\Omega&=& \frac{j}{r_{h}^{2}}\,,\quad 
\phi_e= {\frac{q}{\beta_p^{\frac{1}{2p-1}}}} r_h^{\frac{2(p-1)}{(2p-1)}}\,.
\label{Nonqtes2}
\end{eqnarray}
With these in hand it is easy to check the validity of the standard first law (\ref{first-law}).
As before, one can also consider the {pressure-volume  $P-V$ term, given previously in (\ref{eq:pres-vol}),
as well as vary the coupling constant $\beta_p$, giving raise to the following generalized first law:} 
\be
\delta {\cal M}=T\delta  {\cal S}+\phi_e \delta  {\cal Q}_e+\Omega \delta  {\cal J}+V\delta P+\Pi_{\beta} \delta \beta_p\,, \label{eq:first-law-beta}
\ee
where the new contribution appears due to the parameter $\beta_p$, which reads
\be
\Pi_\beta={-\frac{(-8)^{p} q^{2p}}{16 \beta_p^{\frac{2p}{2p-1}} r_h^{\frac{2(1-p)}{2p-1}}}\left(\frac{p-1}{2p-1}\right)^{2p-1}}\,,
\ee
and, as before, the standard first law (\ref{first-law}) is recovered by fixing $\ell$
and $\beta_p$, respectively. Moreover, by performing a (dimensional) scaling argument:
\begin{eqnarray}
&&r_h\to \alpha r_h\,, \,\, \ell\to \alpha \ell\,,\,\, j\to \alpha j\,,\,\, q\to \alpha^{0} q\,,\nonumber\\
&&\beta_p\to \alpha^{2(p-1)} \beta_p\,, 
\end{eqnarray} 
we obtain that 
\be
2PV=TS+\Omega {\cal J}+2(p-1)\Pi_\beta \beta_p\,.
\ee
Similarly,  
but without rescaling the horizon radius and $\beta_p$, while considering 
$j\to \alpha j, \ell\to \ell/\alpha, q\to \alpha^{1/p}q$,
one finds the following relation:
\begin{eqnarray}
{\cal M}=\frac{1}{2}\left[\Omega {\cal J}+\left(\frac{2p-1}{p}\right)\phi_e {\cal Q}_e+2PV\right].
\label{smarr2}
\end{eqnarray}
Finally, combining these last two expressions, one obtains the following Smarr formula:
\be
    {\cal M}=\Omega {\cal J}+\frac{1}{2}T{\cal S}+\left(\frac{2p-1}{2p}\right)\phi_e{\cal Q}_e+{(p-1)}\Pi_\beta \beta_p\,.
\ee

\subsection{Soliton
as the ground state and Cardy formula}
One of the main goals of our work is to test the validity of a Cardy-like formula for the entropy of black holes 
with various asymptotics
and matter couplings. In order to do so, it is crucial to correctly identify the ground state of the theory, as the Cardy formula relies on computing the degeneracy of excited states above a well-defined vacuum. In our context, the natural candidate for the ground state is a soliton, that is, a completely regular, horizonless solution, devoid of any integration constants. A particularly meaningful choice for such a soliton arises from taking the uncharged, static black hole solution and performing a double Wick rotation. This procedure removes the horizon and yields a smooth geometry, provided appropriate regularity conditions are imposed.

The resulting configuration corresponds precisely to the well-known AdS soliton  \cite{Horowitz:1998ha}, which plays the role of a low-energy reference background in asymptotically AdS spacetimes. Within this framework, the mass of the soliton appears as a negative shift from pure AdS and sets the vacuum energy of the theory, allowing the Cardy-like formula to be applied in a consistent and physically meaningful way. In our case, the soliton of the theory \eqref{action1} is obtained by the following identifications in the metric \eqref{BTZ1} and \eqref{BTZ0}: $j=0=q\,,m=1,$ and:
\be 
t\to i\varphi\,,\quad \varphi \to it\,,\quad \cosh\rho=\frac{r}{\ell}\,,
\ee 
which yields 
\begin{eqnarray}
   &&ds^2=-\ell^2\cosh^2\!\rho dt^2+d\rho^2+\sinh^2\!\rho d\varphi^2,\\
   &&A=0=K\,.\nonumber
\end{eqnarray}
Note that, although upon the above identifications the field $B$ remains non-trivial, it completely decouples, and all the field equations, including constraints \eqref{FJ-nonlinear}-\eqref{algebraic}, are automatically satisfied,  independently of the form of the field $B$. (The same will happen for the Lifshitz solutions described in the next section.)

{The mass of the above soliton }
can be computed using the quasilocal formalism (see Appendix \eqref{quasilocal}). The calculation yields 
\be 
\Delta_{\text{\tiny soliton}} = -\frac{1}{8}\,. 
\ee
Now, it is a matter of checking to see that the Cardy-like formula \eqref{entropy} for the solution with $p=1$ or $p\not=1$ together with $z=1$ will, in both cases, reproduces the expression for the semiclassical entropy, ${\cal S}=\frac{1}{2}\pi r_h$.

\section{Charged Lifshitz rotating solutions and their Cardy entropy}\label{sec:3}

We now consider a slight generalization of the action \eqref{action1} that will allow for the existence of charged Lifshitz rotating solutions. In fact, we will consider two Chern--Simons generalized terms together with the inclusion of a dilaton, $\Phi$, coupled to two gauge fields $A_{\mu}^{(i)}$, with $i\in \{1,2\}$,
\begin{eqnarray}
 S&=&\frac{1}{16\pi}  \int d^3x\sqrt{-g}\Big(R+\frac{z(z+1)}{\ell^2}\nonumber-\frac{1}{2} \partial_\mu \Phi \partial^{\mu} \Phi\nonumber\\
 &&-\sum_{i=1}^{2} e^{\alpha_i \Phi} F^{(i)}_{\mu\nu}F^{(i)\mu\nu}\Big) \nonumber\\
 &&+\sum_{i=1}^{2}\frac{\lambda_i}{4\pi}\int A^{(i)}\wedge H^{(i)}\wedge K^{(i)}\,.\label{eq:action-lifshitz} 
\end{eqnarray}
The equations of motion are given by
\begin{eqnarray}
 &&G_{\mu \nu}-\frac{z(z+1)}{2 \ell^2} g_{\mu \nu}-\left(\frac{1}{2} \partial_\mu \Phi \partial_\nu \Phi-\frac{1}{4} g_{\mu \nu} (\partial_\sigma \Phi \partial^{\sigma}\Phi)\right)\nonumber\\
 &&-\sum_{i=1}^{2} e^{\alpha_i \Phi} \Big( 2  F^{(i)}_{\mu \sigma}F_{\nu}^{(i)\sigma}-\frac{1}{2} g_{\mu \nu} F^{(i)}_{\rho\sigma}F^{(i)\rho\sigma}\Big)=0\,,\\
&&\Box{\Phi}-\sum_{i=1}^{2}\Big( \alpha_i e^{\alpha_i Pphi} F^{(i)}_{\mu\nu}F^{(i)\mu\nu}\Big)=0\,,\\
&&\nabla_\mu \left(e^{\alpha_i \Phi} F^{(i)\mu\nu}\right)=\lambda_i\, *\left(H^{(i)}\wedge K^{(i)}\right)^{\nu}=J^{(i) \nu}\,,\label{eq:J}
\end{eqnarray}

For this model, we can construct charged and rotating Lifshitz black hole solutions within the ansatz \eqref{BTZ1}, namely
\begin{eqnarray}
f(r)&=&\frac{r^2}{\ell^2}+\frac{3-z}{2 (2-z)} \frac{j^2}{r^2}-m\bigg(\frac{\ell}{r}\bigg)^{z-1}\,,\nonumber\\
h(r)&=&\frac{j}{\ell^{z-1}r^{3-z}}\,,\quad  N(r)=\left(\frac{r}{\ell}\right)^{z-1}\,,
\end{eqnarray}
and together with
\begin{eqnarray}
\label{solLi}
\Phi&=&\sqrt{2(z-1)}\log\left(\frac{r}{r_\phi}\right)\,,\nonumber\\
A^{(1)}&=& \sqrt{\frac{z-1}{2(z+1)}}\frac{1}{\ell^z r_\phi}\left(r^{z+1}-r_{h}^{z+1}\right) dt\,,\\
A^{(2)}&=& \displaystyle{\sqrt{\frac{z-1}{4 (2-z)}}\frac{jr_\phi}{\ell^{z-1}}\left(\frac{1}{r^{3-z}}-\frac{1}{r_h^{3-z}}\right) dt.}\nonumber
\end{eqnarray}
Here, we note that the constant of integration of vector field $A^{(1)}$ is fixed to ensure consistency with the desired asymptotic behavior, 
while
\be 
\alpha_1=-\frac{2}{\sqrt{2(z-1)}}=-\alpha_2\,, 
\ee
which is similar to what happened in Ref. \cite{Tarrio:2011de}. As before, $r_h$ is the location of the horizon and $r_\phi$ is an arbitrary scale. In these expressions, $z$ is the well-known Lifshitz dynamical exponent, and in our case $z\in (1,2)$. All of this is possible only if in the field equations \eqref{eq:J}
the $0$-form fields $B^{(i)}=B^{(i)}(t,r)$ and $C^{(i)}=C^{(i)}(r)$ (with $ i \in \{1,2\}$) are subject to the following conditions:
\begin{eqnarray*}
J^{(1)\varphi}&=&-\frac{\sqrt{2(z^2-1)}(3-z) r_\phi j}{2\ell r^4}\,,\label{Jphi1}\\
J^{(2)\varphi}&=&\frac{\sqrt{z-1}(3-z)^2\,j^2}{2\sqrt{2-z} r_\phi r^4}\,,\label{Jphi2}
\end{eqnarray*}
where,  the 0-form fields $B^{(i)}$ and $C^{(i)}$ can be taken in the following way
($ i \in \{1,2\}$):
\begin{eqnarray}
\label{solLi2}
B^{(i)}(t,r)&=&\frac{t}{\lambda_i r}\,, \nonumber\\
C^{(1)}(r)&=&-\frac{\sqrt{2(z-1)(z+1)}(3-z) j r_\phi}{2 (2-z) \ell^z r^{2-z} }\,,\\
C^{(2)}(r)&=&\frac{(3-z)^2 \sqrt{z-1} j^2}{2 (2-z)^{\frac{3}{2}} r_\phi \ell^{z-1}  r^{2-z}}\,.\nonumber
\end{eqnarray}
Note that upon $z\to 1$, we recover the uncharged rotating BTZ solution (together with its thermodynamics -- see below).

\subsection{Thermodynamics}
The extensive thermodynamic quantities associated with this solution are as follows:
\begin{eqnarray}
\label{Lifqtes}
\mathcal{M}&=&\frac{m}{8}=\frac{r_{h}^{z+1}}{8\ell^{z+1}}+\frac{(3-z) j^2}{16(2-z) \ell^{z-1} r_{h}^{3-z} }\,,\nonumber\\
\mathcal{S}&=&\frac{\pi r_h}{2}\,,\quad 
\mathcal{J}=\frac{(3-z) j}{8}\,,\nonumber\\
\mathcal{Q}_e&=&\frac{\sqrt{z-1}\,(3-z) j }{4 \sqrt{2-z} r_\phi}\,,
\end{eqnarray}
while the intensive thermodynamic quantities read
\begin{eqnarray}
T&=&\frac{N(r_h)f'(r_h)}{4\pi}=\frac{(z+1) r_{h}^{z}}{4 \pi \ell^{z+1}}-\frac{(3-z)^2 j^2}{8 \pi (2-z) \ell^{z-1} r_{h}^{4-z} }\,,\nonumber\\
\Omega&=& \frac{j}{\ell^{z-1}r_{h}^{3-z}}\,,\quad 
\phi_e= \sqrt{\frac{z-1}{4(2-z)}}\frac{r_\phi j}{ \ell^{z-1}r_h^{3-z}}\,.
\label{Lifqtes2}
\end{eqnarray}
Note that the charge ${\cal Q}_e$ and the angular momentum ${\cal J}$ are no longer independent, as we have  
\be 
{\cal Q}_e=\sqrt{\left(\frac{z-1}{2-z}\right)}\frac{2{\cal J}}{r_\phi}\,.
\ee 
In other words, we have some kind of a ``{\em constrained}'' black hole. This means that, although one can verify that the first law 
\eqref{first-law} remains valid, it is degenerate and not of full cohomogeneity. 

Similar to the previous AdS case, we can now account for the additional two parameters $\ell$ and $r_\phi$, and include the following terms:
\ba 
V&=& \frac{\pi(z-1)(z-3)\ell^2 j^2+2r_h^4(z+1)(z-2)}{4r_h^{3-z}(z-2)\ell^{z-1}}\,,\nonumber\\
\Pi_{r_\phi}&=&\frac{(z-1)(3-z)j^2}{8(2-z)r_\phi \ell^{z-1}r_h^{3-z}}\,,
\ea 
with the pressure $P$ given in (\ref{eq:pres-vol}), upon which 
the extended first law (\ref{eq:first-law-ext}) holds \footnote{{Note that, similar to was done for example in \cite{Cong:2024pvs}, here we fix the `topological charge' associated with non-trivial Lifshitz asymptotics and only vary the `standard' $U(1)$ charge ${\cal Q}_e$. In principle, one could also include in the first law the variation of the Lifshitz charge, as e.g. done in \cite{Kiritsis:2016rcb}. We leave this issue for future studies.}}.

With this, we can now proceed to find the Smarr relations using the Euler scaling argument. Namely, the following scaling $r_h\to \alpha r_h, r_\phi\to \alpha r_\phi, \ell\to \alpha \ell, j \to \alpha j$ implies the following expression:
\be 
{2PV=T{\cal S}+\Omega {\cal J}+\Pi_{r_\phi} r_\phi\,.}
\ee  
Similarly, the scaling where $\ell$ is fixed, namely, $r_h\to \alpha r_h, r_\phi\to \alpha r_\phi, j\to \alpha^2 j$ allows us to obtain
\be \label{SmarrLifshitz2}
{\cal M}=\frac{1}{z+1} \left(T{\cal S}+2\Omega {\cal J}+\phi_e {\cal Q}_e+\Pi_{r_\phi}r_\phi\right),
\ee 
and combining these two equations, we obtain the following Smarr relation:
\be 
{\cal M}=\frac{1}{z+1}(\Omega {\cal J}+\phi_e {\cal Q}_e+2PV)\,.
\label{smarr3}
\ee
At the same time, 
we also have the following ``{\em accidental}''  (degeneracy) relations:
\be 
{\Pi_{r_\phi}r_\phi=\phi_e {\cal Q}_e=\left(\frac{z-1}{2-z}\right)\Omega {\cal J}\,},
\ee 
which can be used to further simplify the above Smarr relations. For example, the Smarr expression \eqref{SmarrLifshitz2} can also be written as:
\begin{eqnarray} 
{\cal M}&=&\frac{1}{z+1}\Big(T{\cal S}+2\Omega {\cal J}+2\phi_e {\cal Q}_e\Big)\nonumber\\
&=&\frac{1}{z+1}\left[T{\cal S}+\left(\frac{2}{2-z}\right)\Omega {\cal J}\right].
\end{eqnarray}

\subsection{Soliton and Cardy formula}

The uncharged non-rotating soliton configuration is obtained, as before, via a double Wick rotation  from the static and uncharged   solution, 
see e.g. Ref. \cite{Ayon-Beato:2015jga} formulas (16)-(20). It reads:
\begin{eqnarray}
\label{sol1}
ds^2&=&-\frac{r^2}{\ell^2}dt^2+\frac{dr^2}{f(r)}+N^2(r)f(r) d \varphi^2,\\
f(r)&=&\frac{r^2}{\ell^2}-\frac{\mu \ell^{z-1}}{r^{z-1}},\quad N(r)=\left(\frac{r}{\ell}\right)^{z-1},\nonumber\\
\mu&=&\left(\frac{2}{z+1}\right)^{\frac{z+1}{z}},
\end{eqnarray}
By means of the quasilocal formulation \cite{Kim:2013zha,Gim:2014nba}, we obtain that the soliton mass $\Delta_{\mbox{\tiny{soliton}}}$ reads (see Appendix \eqref{quasilocal} for details)
\begin{equation}
\label{massol1}
\Delta_{\mbox{\tiny{soliton}}}=-\frac{z}{8} \left(\frac{2}{z+1}\right)^{\frac{z+1}{z}}.
\end{equation}
As in the previous cases, one can easily check that plugging this soliton mass as well as the Lifshitz thermodynamical quantities \eqref{Lifqtes} and \eqref{Lifqtes2} into the Lifshitz Cardy formula \eqref{entropy} with $p=1$, the expression of the entropy ${\cal S}=\frac{\pi r_h}{2}$ is again recovered.

\section{Generalized 
Smarr and Cardy formulas}\label{sec:4}
A quick look at the three formulas \eqref{smarr1}, \eqref{smarr2} and \eqref{smarr3} in the cases of AdS with $z=1$ or Lifshitz with $z\not=1$, and for Maxwell electrodynamics with $p=1$ or nonlinear electrodynamics with $p\not=1$, suggests {the following generalization of the Smarr formula:}
\begin{eqnarray}
{\cal M}=\frac{1}{z+1}\left[\Omega{\cal J}+\left(\frac{2p-1}{p}\right)\phi {\cal Q}_e+2PV\right]\,, 
    \label{genericSmarr}
\end{eqnarray}
from where one can get that 
\be
PV+\frac{(z-1)}{z}{\cal M}={\cal M}- \frac{1}{2}\Omega {\cal J} - \left(\frac{2p - 1}{2p}\right) \phi {\cal Q}_e\,.
\label{pvrel}
\ee
It is interesting to note that for an arbitrary value of $z$ the right-hand side of \eqref{pvrel} is always proportional to the horizon area to the power $z+1$, namely 
\be
PV+\frac{(z-1)}{z}{\cal M}\propto r_h^{z+1}.
\ee
Based on this last observation, {we propose that} a generic expression for the entropy can be obtained via the following generalized Cardy-like formula involving the soliton mass  $\Delta_{\text{\tiny soliton}}$: 
\begin{widetext}
\begin{eqnarray}
 \boxed{ {\cal S} =
2\pi \ell (z+1) \left(-\frac{\Delta_{\text{\tiny soliton}}}{z}\right)^{\frac{z}{z+1}}
\left[
{\cal M} - \frac{1}{2}\Omega {\cal J} - \left(\frac{2p - 1}{2p}\right) \phi_e {\cal Q}_e
\right]^{\frac{1}{z+1}}
}.
\label{entropy}
\end{eqnarray}
\end{widetext}
Note that in the uncharged and static case ${\cal J}=0={\cal Q}_e$, this formula reduces to the one derived in \cite{Gonzalez:2011nz} where the authors made 
made use of the isomorphism between the Lifshitz algebras with dynamical exponents $z$ and $z^{-1}$, and where the soliton was identified as the ground state. 

In the AdS case $z=1$ and in the absence of electric charge, the generic formula \eqref{entropy} can be identified with the 
Cardy formula in the following way. 
Starting from the
standard Cardy formula
\be
S=2\pi\sqrt{\frac{c}{6}\left(L_0-\frac{c}{24}\right)}+2\pi\sqrt{\frac{\bar{c}}{6}\left(\bar{L}_0-\frac{\bar{c}}{24}\right)}\,,
\ee
where $c$ and $\bar{c}$ are the central charges and the energy levels $L_0$ and $\bar{L}_0$ correspond to
\be
L_0=\frac{1}{2}\left({\cal M}\ell+{\cal J}\right),\qquad \bar{L}_0=\frac{1}{2}\left({\cal M}\ell-{\cal J}\right)\,.
\ee
At high energy ${\cal M}\ell\gg\frac{c}{24}$, and in the BTZ case $c=\bar{c}=\frac{3\ell}{2}$, the standard Cardy formula reduces to 
\be 
S\sim 2\pi\sqrt{\frac{\ell}{8}}\, \Big[\sqrt{{\cal M}\ell+{\cal J}}+\sqrt{{\cal M}\ell-{\cal J}}\Big].
\ee
On the other hand, since 
\be 
\sqrt{{\cal M}\ell+{\cal J}}+\sqrt{{\cal M}\ell-{\cal J}}=2\sqrt{\ell}\sqrt{{\cal M} - \frac{1}{2}\Omega {\cal J}}\,,
\ee
upon using the uncharged BTZ thermodynamic quantities \eqref{charges}
it is easy to see that the standard Cardy formula reduces to the expression  \eqref{entropy} with $\Delta_{\text{\tiny soliton}}=-\frac{1}{8}$. Thus, although the entropy formula we propose does not stem from a microscopic state counting in the usual sense, it coincides, at least formally, with the standard Cardy formula in the uncharged AdS case. However, beyond this specific regime, particularly in the presence of electric charge or non-AdS asymptotics, such an interpretation is no longer available, and the formula cannot be directly linked to the eigenvalues of the conformal generators.

\section{Concluding remarks}\label{sec:5}

By analyzing the thermodynamics of a wide variety of rotating and charged black hole systems (constructed in this paper) we have proposed a generalized Cardy-like formula that correctly reproduces their semiclassical entropy. Such a formula, rather than being based on central charges, stems from identifying the mass of the corresponding soliton solution (ground state) obtained via a double Wick rotation of the static uncharged black hole in a given theory.

Our findings suggest a versatile Cardy-type expression that unifies the description of black hole entropy across various setups, encompassing both linear and non-linear electrodynamics, as well as AdS and Lifshitz boundary conditions. Note, however, that although we possess a Cardy-like formula that successfully reproduces the entropy across all configurations studied, the derivation of this formula does not follow from the standard route based on a microscopic counting of energy levels in a two-dimensional conformal field theory. In particular, in the generic case, there is no direct interpretation of the parameters $L_0$ and $\bar{L}_0$ as eigenvalues of the Hamiltonian and angular momentum operators. Nonetheless, the persistence of the Cardy-like entropy formula across such diverse settings strongly suggests a deeper structure underlying these results.

In this context, it is tempting to conjecture that the success of the Cardy-like formula might be 
intimately related to the existence of generalized Smarr relations, as we have 
outlined
here. These identities, which express the black hole mass as a bilinear combination of conserved charges and their thermodynamic potentials, can be viewed as integral versions of the first law of black hole thermodynamics. Our results indicate that the entropy formula we employ, though not derived via a direct counting of microstates, encodes similar structural information to that contained in Smarr-type relations.

In addition, it would be highly desirable to construct and study a broader class of models admitting rotating and electrically charged black hole solutions. This would allow for a more stringent test of the robustness and generality of the proposed Cardy-like entropy formula, especially in cases where the asymptotic structure or the matter content significantly deviates from the canonical AdS-Maxwell setup. Such an investigation would help to uncover the precise conditions under which the formula remains valid, as well as may offer insights into its possible microscopic interpretation.

\acknowledgments

M.B. is supported by proyecto interno UCM-IN-25202. A.C. is partially supported by FONDECYT grant 1250318 and by the GA{\v C}R 22-14791S grant from the Czech Science Foundation.
D.K. is grateful for support from
GA{\v C}R 23-07457S grant of the Czech Science Foundation and the Charles University Research
Center Grant No. UNCE24/SCI/016.
M.H. gratefully acknowledges the University of Paris-Saclay for its warm hospitality during the development of this project.

\appendix


\section{Quasilocal mass of the soliton \label{quasilocal}}

To ensure completeness, the mass of the different solitons will be computed using the quasi-local formalism as introduced in Refs. \cite{Kim:2013zha, Gim:2014nba}. The charge associated to the Killing field $\xi$ is given by 
\begin{equation}
\label{chargequasi} Q(\xi)\!=\!\int_{\cal B}\!
dx_{\mu\nu}\Big(\delta K^{\mu\nu}(\xi)-2\xi^{[\mu} \!\! \int^1_0ds~
\Theta^{\nu]}\Big).
\end{equation}
Here, $K^{\mu\nu}$ corresponds to the Noether potential and $\Theta^{\mu}$ is the boundary term, whose expressions in our case are generically given by 
\begin{eqnarray}\label{eq:thetamu}
\Theta^{\mu}&=&2\sqrt{-g}\biggl[\left(\frac{\partial \mathcal{L}}{\partial R_{\mu\alpha\beta\gamma}}\right)\nabla_{\gamma}\delta g_{\alpha\beta}-\delta g_{\alpha\beta} \nabla_{\gamma}\left(\frac{\partial \mathcal{L}}{\partial R_{\mu\alpha\beta\gamma}}\right)\nonumber\\
&+&\frac{\partial \mathcal{L}}{2\,\partial (\partial_\mu \phi)}\delta \phi+\frac{\partial \mathcal{L}}{2\,\partial (\partial_\mu A_\nu)}\delta A_{\nu}\biggr],
\end{eqnarray}
and 
\begin{eqnarray}\label{eq:kmunu}
K^{\mu\nu}&=&\sqrt{-g}\Bigg[2\left(\frac{\partial \mathcal{L}}{\partial R_{\mu\nu\rho\sigma}}\right)\nabla_{\rho} \xi_{\sigma} - 4\xi_{\sigma}\nabla_{\rho} \left(\frac{\partial \mathcal{L}}{\partial R_{\mu\nu\rho\sigma}}\right) \nonumber\\
&-&\frac{\partial \mathcal{L}}{2\,\partial (\partial_\mu A_\nu)} \xi^{\rho} A_{\rho}\Bigg].
\end{eqnarray}
The integration variable $s$ in \eqref{chargequasi}  is a parameter that allows the interpolation between the solution of interest at $s = 1$ and the asymptotic configuration at $s = 0$. In addition,  $\delta K^{\mu\nu}(\xi) =
K^{\mu\nu}_{s=1}(\xi)-K^{\mu\nu}_{s=0}(\xi)$ stands for the difference of the Noether potential between these two configurations, while $dx_{\mu\nu}$ represents the integration over the two-dimensional boundary ${\cal B}$. It is interesting to note that the Chern--Simons terms of the {actions \eqref{action1} and \eqref{eq:action-lifshitz} considered in the main text do} not contribute to the charges associated with any Killing field.

Let us apply these formulas for the different soliton solutions discussed in the main text.
\begin{enumerate}[label=(\roman*)]
    \item In the case of the AdS soliton, we have 
    \be 
    \delta K^{rt}=-\frac{1}{8 \pi \ell}\,, \quad \int_0^1 ds \Theta^r=\frac{1}{16 \pi \ell}\,,
    \ee
and, hence, the mass of the AdS soliton is given by $\Delta_{\mbox{\tiny{AdSsoliton}}}=-\frac{1}{8}.$
    
\item For the static Lifshitz soliton \eqref{sol1}, 
\be 
\int_0^1 ds \Theta^r=-\frac{(z-2) \mu }{16 \pi \ell },  \quad \delta K^{rt}=-\frac{\mu}{8 \pi \ell}\,,
\ee
and the Lifshitz soliton mass is
\be 
\Delta_{\mbox{\tiny{soliton}}}=-\frac{z\mu}{8}=-\frac{z}{8} \left(\frac{2}{z+1}\right)^{\frac{z+1}{z}}\,.
\ee
\end{enumerate}


\begin{thebibliography}{25}%
\makeatletter
\providecommand \@ifxundefined [1]{%
 \@ifx{#1\undefined}
}%
\providecommand \@ifnum [1]{%
 \ifnum #1\expandafter \@firstoftwo
 \else \expandafter \@secondoftwo
 \fi
}%
\providecommand \@ifx [1]{%
 \ifx #1\expandafter \@firstoftwo
 \else \expandafter \@secondoftwo
 \fi
}%
\providecommand \natexlab [1]{#1}%
\providecommand \enquote  [1]{``#1''}%
\providecommand \bibnamefont  [1]{#1}%
\providecommand \bibfnamefont [1]{#1}%
\providecommand \citenamefont [1]{#1}%
\providecommand \href@noop [0]{\@secondoftwo}%
\providecommand \href [0]{\begingroup \@sanitize@url \@href}%
\providecommand \@href[1]{\@@startlink{#1}\@@href}%
\providecommand \@@href[1]{\endgroup#1\@@endlink}%
\providecommand \@sanitize@url [0]{\catcode `\\12\catcode `\$12\catcode `\&12\catcode `\#12\catcode `\^12\catcode `\_12\catcode `\%12\relax}%
\providecommand \@@startlink[1]{}%
\providecommand \@@endlink[0]{}%
\providecommand \url  [0]{\begingroup\@sanitize@url \@url }%
\providecommand \@url [1]{\endgroup\@href {#1}{\urlprefix }}%
\providecommand \urlprefix  [0]{URL }%
\providecommand \Eprint [0]{\href }%
\providecommand \doibase [0]{https://doi.org/}%
\providecommand \selectlanguage [0]{\@gobble}%
\providecommand \bibinfo  [0]{\@secondoftwo}%
\providecommand \bibfield  [0]{\@secondoftwo}%
\providecommand \translation [1]{[#1]}%
\providecommand \BibitemOpen [0]{}%
\providecommand \bibitemStop [0]{}%
\providecommand \bibitemNoStop [0]{.\EOS\space}%
\providecommand \EOS [0]{\spacefactor3000\relax}%
\providecommand \BibitemShut  [1]{\csname bibitem#1\endcsname}%
\let\auto@bib@innerbib\@empty
\bibitem [{\citenamefont {Maldacena}(1998)}]{Maldacena:1997re}%
  \BibitemOpen
  \bibfield  {author} {\bibinfo {author} {\bibfnamefont {J.~M.}\ \bibnamefont {Maldacena}},\ }\bibfield  {title} {\bibinfo {title} {{The Large $N$ limit of superconformal field theories and supergravity}},\ }\href {https://doi.org/10.4310/ATMP.1998.v2.n2.a1} {\bibfield  {journal} {\bibinfo  {journal} {Adv. Theor. Math. Phys.}\ }\textbf {\bibinfo {volume} {2}},\ \bibinfo {pages} {231} (\bibinfo {year} {1998})},\ \Eprint {https://arxiv.org/abs/hep-th/9711200} {arXiv:hep-th/9711200} \BibitemShut {NoStop}%
\bibitem [{\citenamefont {Brown}\ and\ \citenamefont {Henneaux}(1986)}]{Brown:1986nw}%
  \BibitemOpen
  \bibfield  {author} {\bibinfo {author} {\bibfnamefont {J.~D.}\ \bibnamefont {Brown}}\ and\ \bibinfo {author} {\bibfnamefont {M.}~\bibnamefont {Henneaux}},\ }\bibfield  {title} {\bibinfo {title} {{Central Charges in the Canonical Realization of Asymptotic Symmetries: An Example from Three-Dimensional Gravity}},\ }\href {https://doi.org/10.1007/BF01211590} {\bibfield  {journal} {\bibinfo  {journal} {Commun. Math. Phys.}\ }\textbf {\bibinfo {volume} {104}},\ \bibinfo {pages} {207} (\bibinfo {year} {1986})}\BibitemShut {NoStop}%
\bibitem [{\citenamefont {Banados}\ \emph {et~al.}(1992)\citenamefont {Banados}, \citenamefont {Teitelboim},\ and\ \citenamefont {Zanelli}}]{Banados:1992wn}%
  \BibitemOpen
  \bibfield  {author} {\bibinfo {author} {\bibfnamefont {M.}~\bibnamefont {Banados}}, \bibinfo {author} {\bibfnamefont {C.}~\bibnamefont {Teitelboim}},\ and\ \bibinfo {author} {\bibfnamefont {J.}~\bibnamefont {Zanelli}},\ }\bibfield  {title} {\bibinfo {title} {{The Black hole in three-dimensional space-time}},\ }\href {https://doi.org/10.1103/PhysRevLett.69.1849} {\bibfield  {journal} {\bibinfo  {journal} {Phys. Rev. Lett.}\ }\textbf {\bibinfo {volume} {69}},\ \bibinfo {pages} {1849} (\bibinfo {year} {1992})},\ \Eprint {https://arxiv.org/abs/hep-th/9204099} {arXiv:hep-th/9204099} \BibitemShut {NoStop}%
\bibitem [{\citenamefont {Strominger}(1998)}]{Strominger:1997eq}%
  \BibitemOpen
  \bibfield  {author} {\bibinfo {author} {\bibfnamefont {A.}~\bibnamefont {Strominger}},\ }\bibfield  {title} {\bibinfo {title} {{Black hole entropy from near horizon microstates}},\ }\href {https://doi.org/10.1088/1126-6708/1998/02/009} {\bibfield  {journal} {\bibinfo  {journal} {JHEP}\ }\textbf {\bibinfo {volume} {02}},\ \bibinfo {pages} {009}},\ \Eprint {https://arxiv.org/abs/hep-th/9712251} {arXiv:hep-th/9712251} \BibitemShut {NoStop}%
\bibitem [{\citenamefont {Cardy}(1986)}]{Cardy:1986ie}%
  \BibitemOpen
  \bibfield  {author} {\bibinfo {author} {\bibfnamefont {J.~L.}\ \bibnamefont {Cardy}},\ }\bibfield  {title} {\bibinfo {title} {{Operator Content of Two-Dimensional Conformally Invariant Theories}},\ }\href {https://doi.org/10.1016/0550-3213(86)90552-3} {\bibfield  {journal} {\bibinfo  {journal} {Nucl. Phys. B}\ }\textbf {\bibinfo {volume} {270}},\ \bibinfo {pages} {186} (\bibinfo {year} {1986})}\BibitemShut {NoStop}%
\bibitem [{\citenamefont {Verlinde}(2000)}]{Verlinde:2000wg}%
  \BibitemOpen
  \bibfield  {author} {\bibinfo {author} {\bibfnamefont {E.~P.}\ \bibnamefont {Verlinde}},\ }\bibfield  {title} {\bibinfo {title} {{On the holographic principle in a radiation dominated universe}},\ }\href@noop {} {\  (\bibinfo {year} {2000})},\ \Eprint {https://arxiv.org/abs/hep-th/0008140} {arXiv:hep-th/0008140} \BibitemShut {NoStop}%
\bibitem [{\citenamefont {Correa}\ \emph {et~al.}(2011)\citenamefont {Correa}, \citenamefont {Martinez},\ and\ \citenamefont {Troncoso}}]{Correa:2010hf}%
  \BibitemOpen
  \bibfield  {author} {\bibinfo {author} {\bibfnamefont {F.}~\bibnamefont {Correa}}, \bibinfo {author} {\bibfnamefont {C.}~\bibnamefont {Martinez}},\ and\ \bibinfo {author} {\bibfnamefont {R.}~\bibnamefont {Troncoso}},\ }\bibfield  {title} {\bibinfo {title} {{Scalar solitons and the microscopic entropy of hairy black holes in three dimensions}},\ }\href {https://doi.org/10.1007/JHEP01(2011)034} {\bibfield  {journal} {\bibinfo  {journal} {JHEP}\ }\textbf {\bibinfo {volume} {01}},\ \bibinfo {pages} {034}},\ \Eprint {https://arxiv.org/abs/1010.1259} {arXiv:1010.1259 [hep-th]} \BibitemShut {NoStop}%
\bibitem [{\citenamefont {Gonzalez}\ \emph {et~al.}(2011)\citenamefont {Gonzalez}, \citenamefont {Tempo},\ and\ \citenamefont {Troncoso}}]{Gonzalez:2011nz}%
  \BibitemOpen
  \bibfield  {author} {\bibinfo {author} {\bibfnamefont {H.~A.}\ \bibnamefont {Gonzalez}}, \bibinfo {author} {\bibfnamefont {D.}~\bibnamefont {Tempo}},\ and\ \bibinfo {author} {\bibfnamefont {R.}~\bibnamefont {Troncoso}},\ }\bibfield  {title} {\bibinfo {title} {{Field theories with anisotropic scaling in 2D, solitons and the microscopic entropy of asymptotically Lifshitz black holes}},\ }\href {https://doi.org/10.1007/JHEP11(2011)066} {\bibfield  {journal} {\bibinfo  {journal} {JHEP}\ }\textbf {\bibinfo {volume} {11}},\ \bibinfo {pages} {066}},\ \Eprint {https://arxiv.org/abs/1107.3647} {arXiv:1107.3647 [hep-th]} \BibitemShut {NoStop}%
\bibitem [{\citenamefont {Bravo~Gaete}\ \emph {et~al.}(2017)\citenamefont {Bravo~Gaete}, \citenamefont {Guajardo},\ and\ \citenamefont {Hassaine}}]{BravoGaete:2017dso}%
  \BibitemOpen
  \bibfield  {author} {\bibinfo {author} {\bibfnamefont {M.}~\bibnamefont {Bravo~Gaete}}, \bibinfo {author} {\bibfnamefont {L.}~\bibnamefont {Guajardo}},\ and\ \bibinfo {author} {\bibfnamefont {M.}~\bibnamefont {Hassaine}},\ }\bibfield  {title} {\bibinfo {title} {{A Cardy-like formula for rotating black holes with planar horizon}},\ }\href {https://doi.org/10.1007/JHEP04(2017)092} {\bibfield  {journal} {\bibinfo  {journal} {JHEP}\ }\textbf {\bibinfo {volume} {04}},\ \bibinfo {pages} {092}},\ \Eprint {https://arxiv.org/abs/1702.02416} {arXiv:1702.02416 [hep-th]} \BibitemShut {NoStop}%
\bibitem [{\citenamefont {Alkac}\ \emph {et~al.}(2025)\citenamefont {Alkac}, \citenamefont {Guajardo},\ and\ \citenamefont {Ozsahin}}]{Alkac:2024hvu}%
  \BibitemOpen
  \bibfield  {author} {\bibinfo {author} {\bibfnamefont {G.}~\bibnamefont {Alkac}}, \bibinfo {author} {\bibfnamefont {L.}~\bibnamefont {Guajardo}},\ and\ \bibinfo {author} {\bibfnamefont {H.}~\bibnamefont {Ozsahin}},\ }\bibfield  {title} {\bibinfo {title} {{Microscopic entropy of static black holes in 3D Lovelock gravities}},\ }\href {https://doi.org/10.1103/PhysRevD.111.044006} {\bibfield  {journal} {\bibinfo  {journal} {Phys. Rev. D}\ }\textbf {\bibinfo {volume} {111}},\ \bibinfo {pages} {044006} (\bibinfo {year} {2025})},\ \Eprint {https://arxiv.org/abs/2409.03865} {arXiv:2409.03865 [hep-th]} \BibitemShut {NoStop}%
\bibitem [{\citenamefont {Ay\'on-Beato}\ \emph {et~al.}(2015)\citenamefont {Ay\'on-Beato}, \citenamefont {Bravo-Gaete}, \citenamefont {Correa}, \citenamefont {Hassa\"\i{}ne}, \citenamefont {Ju\'arez-Aubry},\ and\ \citenamefont {Oliva}}]{Ayon-Beato:2015jga}%
  \BibitemOpen
  \bibfield  {author} {\bibinfo {author} {\bibfnamefont {E.}~\bibnamefont {Ay\'on-Beato}}, \bibinfo {author} {\bibfnamefont {M.}~\bibnamefont {Bravo-Gaete}}, \bibinfo {author} {\bibfnamefont {F.}~\bibnamefont {Correa}}, \bibinfo {author} {\bibfnamefont {M.}~\bibnamefont {Hassa\"\i{}ne}}, \bibinfo {author} {\bibfnamefont {M.~M.}\ \bibnamefont {Ju\'arez-Aubry}},\ and\ \bibinfo {author} {\bibfnamefont {J.}~\bibnamefont {Oliva}},\ }\bibfield  {title} {\bibinfo {title} {{First law and anisotropic Cardy formula for three-dimensional Lifshitz black holes}},\ }\href {https://doi.org/10.1103/PhysRevD.91.064006} {\bibfield  {journal} {\bibinfo  {journal} {Phys. Rev. D}\ }\textbf {\bibinfo {volume} {91}},\ \bibinfo {pages} {064006} (\bibinfo {year} {2015})},\ \bibinfo {note} {[Addendum: Phys.Rev.D 96, 049903 (2017)]},\ \Eprint {https://arxiv.org/abs/1501.01244} {arXiv:1501.01244 [gr-qc]} \BibitemShut {NoStop}%
\bibitem [{\citenamefont {Ay\'on-Beato}\ \emph {et~al.}(2019)\citenamefont {Ay\'on-Beato}, \citenamefont {Bravo-Gaete}, \citenamefont {Correa}, \citenamefont {Hassaine},\ and\ \citenamefont {Ju\'arez-Aubry}}]{Ayon-Beato:2019kmz}%
  \BibitemOpen
  \bibfield  {author} {\bibinfo {author} {\bibfnamefont {E.}~\bibnamefont {Ay\'on-Beato}}, \bibinfo {author} {\bibfnamefont {M.}~\bibnamefont {Bravo-Gaete}}, \bibinfo {author} {\bibfnamefont {F.}~\bibnamefont {Correa}}, \bibinfo {author} {\bibfnamefont {M.}~\bibnamefont {Hassaine}},\ and\ \bibinfo {author} {\bibfnamefont {M.~M.}\ \bibnamefont {Ju\'arez-Aubry}},\ }\bibfield  {title} {\bibinfo {title} {{Microscopic entropy of higher-dimensional nonminimally dressed Lifshitz black holes}},\ }\href {https://doi.org/10.1103/PhysRevD.100.044024} {\bibfield  {journal} {\bibinfo  {journal} {Phys. Rev. D}\ }\textbf {\bibinfo {volume} {100}},\ \bibinfo {pages} {044024} (\bibinfo {year} {2019})},\ \Eprint {https://arxiv.org/abs/1904.09391} {arXiv:1904.09391 [hep-th]} \BibitemShut {NoStop}%
\bibitem [{\citenamefont {Bravo-Gaete}\ \emph {et~al.}(2015)\citenamefont {Bravo-Gaete}, \citenamefont {Gomez},\ and\ \citenamefont {Hassaine}}]{Bravo-Gaete:2015wua}%
  \BibitemOpen
  \bibfield  {author} {\bibinfo {author} {\bibfnamefont {M.}~\bibnamefont {Bravo-Gaete}}, \bibinfo {author} {\bibfnamefont {S.}~\bibnamefont {Gomez}},\ and\ \bibinfo {author} {\bibfnamefont {M.}~\bibnamefont {Hassaine}},\ }\bibfield  {title} {\bibinfo {title} {{Towards the Cardy formula for hyperscaling violation black holes}},\ }\href {https://doi.org/10.1103/PhysRevD.91.124038} {\bibfield  {journal} {\bibinfo  {journal} {Phys. Rev. D}\ }\textbf {\bibinfo {volume} {91}},\ \bibinfo {pages} {124038} (\bibinfo {year} {2015})},\ \Eprint {https://arxiv.org/abs/1505.00702} {arXiv:1505.00702 [hep-th]} \BibitemShut {NoStop}%
\bibitem [{\citenamefont {Hassaine}\ and\ \citenamefont {Martinez}(2007)}]{Hassaine:2007py}%
  \BibitemOpen
  \bibfield  {author} {\bibinfo {author} {\bibfnamefont {M.}~\bibnamefont {Hassaine}}\ and\ \bibinfo {author} {\bibfnamefont {C.}~\bibnamefont {Martinez}},\ }\bibfield  {title} {\bibinfo {title} {{Higher-dimensional black holes with a conformally invariant Maxwell source}},\ }\href {https://doi.org/10.1103/PhysRevD.75.027502} {\bibfield  {journal} {\bibinfo  {journal} {Phys. Rev. D}\ }\textbf {\bibinfo {volume} {75}},\ \bibinfo {pages} {027502} (\bibinfo {year} {2007})},\ \Eprint {https://arxiv.org/abs/hep-th/0701058} {arXiv:hep-th/0701058} \BibitemShut {NoStop}%
\bibitem [{\citenamefont {Hassaine}\ and\ \citenamefont {Martinez}(2008)}]{Hassaine:2008pw}%
  \BibitemOpen
  \bibfield  {author} {\bibinfo {author} {\bibfnamefont {M.}~\bibnamefont {Hassaine}}\ and\ \bibinfo {author} {\bibfnamefont {C.}~\bibnamefont {Martinez}},\ }\bibfield  {title} {\bibinfo {title} {{Higher-dimensional charged black holes solutions with a nonlinear electrodynamics source}},\ }\href {https://doi.org/10.1088/0264-9381/25/19/195023} {\bibfield  {journal} {\bibinfo  {journal} {Class. Quant. Grav.}\ }\textbf {\bibinfo {volume} {25}},\ \bibinfo {pages} {195023} (\bibinfo {year} {2008})},\ \Eprint {https://arxiv.org/abs/0803.2946} {arXiv:0803.2946 [hep-th]} \BibitemShut {NoStop}%
\bibitem [{\citenamefont {Deshpande}\ and\ \citenamefont {Lunin}(2024)}]{Deshpande:2024vbn}%
  \BibitemOpen
  \bibfield  {author} {\bibinfo {author} {\bibfnamefont {R.}~\bibnamefont {Deshpande}}\ and\ \bibinfo {author} {\bibfnamefont {O.}~\bibnamefont {Lunin}},\ }\bibfield  {title} {\bibinfo {title} {{Rotating Einstein-Maxwell black holes in odd dimensions}},\ }\href@noop {} {\  (\bibinfo {year} {2024})},\ \Eprint {https://arxiv.org/abs/2411.01795} {arXiv:2411.01795 [hep-th]} \BibitemShut {NoStop}%
\bibitem [{\citenamefont {Hale}\ \emph {et~al.}(2024)\citenamefont {Hale}, \citenamefont {Hull}, \citenamefont {Kubiz\v{n}\'ak}, \citenamefont {Mann},\ and\ \citenamefont {Men\v{s}\'\i{}kov\'a}}]{Hale:2024zvu}%
  \BibitemOpen
  \bibfield  {author} {\bibinfo {author} {\bibfnamefont {T.}~\bibnamefont {Hale}}, \bibinfo {author} {\bibfnamefont {B.~R.}\ \bibnamefont {Hull}}, \bibinfo {author} {\bibfnamefont {D.}~\bibnamefont {Kubiz\v{n}\'ak}}, \bibinfo {author} {\bibfnamefont {R.~B.}\ \bibnamefont {Mann}},\ and\ \bibinfo {author} {\bibfnamefont {J.}~\bibnamefont {Men\v{s}\'\i{}kov\'a}},\ }\bibfield  {title} {\bibinfo {title} {{New interpretation of the original charged BTZ black hole spacetime}},\ }\href@noop {} {\  (\bibinfo {year} {2024})},\ \Eprint {https://arxiv.org/abs/2412.04329} {arXiv:2412.04329 [gr-qc]} \BibitemShut {NoStop}%
\bibitem [{\citenamefont {Cardenas}\ \emph {et~al.}(2014)\citenamefont {Cardenas}, \citenamefont {Fuentealba},\ and\ \citenamefont {Mart\'\i{}nez}}]{Cardenas:2014kaa}%
  \BibitemOpen
  \bibfield  {author} {\bibinfo {author} {\bibfnamefont {M.}~\bibnamefont {Cardenas}}, \bibinfo {author} {\bibfnamefont {O.}~\bibnamefont {Fuentealba}},\ and\ \bibinfo {author} {\bibfnamefont {C.}~\bibnamefont {Mart\'\i{}nez}},\ }\bibfield  {title} {\bibinfo {title} {{Three-dimensional black holes with conformally coupled scalar and gauge fields}},\ }\href {https://doi.org/10.1103/PhysRevD.90.124072} {\bibfield  {journal} {\bibinfo  {journal} {Phys. Rev. D}\ }\textbf {\bibinfo {volume} {90}},\ \bibinfo {pages} {124072} (\bibinfo {year} {2014})},\ \Eprint {https://arxiv.org/abs/1408.1401} {arXiv:1408.1401 [hep-th]} \BibitemShut {NoStop}%
\bibitem [{\citenamefont {Gonzalez}\ \emph {et~al.}(2009)\citenamefont {Gonzalez}, \citenamefont {Hassaine},\ and\ \citenamefont {Martinez}}]{Gonzalez:2009nn}%
  \BibitemOpen
  \bibfield  {author} {\bibinfo {author} {\bibfnamefont {H.~A.}\ \bibnamefont {Gonzalez}}, \bibinfo {author} {\bibfnamefont {M.}~\bibnamefont {Hassaine}},\ and\ \bibinfo {author} {\bibfnamefont {C.}~\bibnamefont {Martinez}},\ }\bibfield  {title} {\bibinfo {title} {{Thermodynamics of charged black holes with a nonlinear electrodynamics source}},\ }\href {https://doi.org/10.1103/PhysRevD.80.104008} {\bibfield  {journal} {\bibinfo  {journal} {Phys. Rev. D}\ }\textbf {\bibinfo {volume} {80}},\ \bibinfo {pages} {104008} (\bibinfo {year} {2009})},\ \Eprint {https://arxiv.org/abs/0909.1365} {arXiv:0909.1365 [hep-th]} \BibitemShut {NoStop}%
\bibitem [{\citenamefont {Horowitz}\ and\ \citenamefont {Myers}(1998)}]{Horowitz:1998ha}%
  \BibitemOpen
  \bibfield  {author} {\bibinfo {author} {\bibfnamefont {G.~T.}\ \bibnamefont {Horowitz}}\ and\ \bibinfo {author} {\bibfnamefont {R.~C.}\ \bibnamefont {Myers}},\ }\bibfield  {title} {\bibinfo {title} {{The AdS / CFT correspondence and a new positive energy conjecture for general relativity}},\ }\href {https://doi.org/10.1103/PhysRevD.59.026005} {\bibfield  {journal} {\bibinfo  {journal} {Phys. Rev. D}\ }\textbf {\bibinfo {volume} {59}},\ \bibinfo {pages} {026005} (\bibinfo {year} {1998})},\ \Eprint {https://arxiv.org/abs/hep-th/9808079} {arXiv:hep-th/9808079} \BibitemShut {NoStop}%
\bibitem [{\citenamefont {Tarrio}\ and\ \citenamefont {Vandoren}(2011)}]{Tarrio:2011de}%
  \BibitemOpen
  \bibfield  {author} {\bibinfo {author} {\bibfnamefont {J.}~\bibnamefont {Tarrio}}\ and\ \bibinfo {author} {\bibfnamefont {S.}~\bibnamefont {Vandoren}},\ }\bibfield  {title} {\bibinfo {title} {{Black holes and black branes in Lifshitz spacetimes}},\ }\href {https://doi.org/10.1007/JHEP09(2011)017} {\bibfield  {journal} {\bibinfo  {journal} {JHEP}\ }\textbf {\bibinfo {volume} {09}},\ \bibinfo {pages} {017}},\ \Eprint {https://arxiv.org/abs/1105.6335} {arXiv:1105.6335 [hep-th]} \BibitemShut {NoStop}%
\bibitem [{\citenamefont {Cong}\ \emph {et~al.}(2024)\citenamefont {Cong}, \citenamefont {Kubiz\v{n}\'ak}, \citenamefont {Mann},\ and\ \citenamefont {Visser}}]{Cong:2024pvs}%
  \BibitemOpen
  \bibfield  {author} {\bibinfo {author} {\bibfnamefont {W.}~\bibnamefont {Cong}}, \bibinfo {author} {\bibfnamefont {D.}~\bibnamefont {Kubiz\v{n}\'ak}}, \bibinfo {author} {\bibfnamefont {R.~B.}\ \bibnamefont {Mann}},\ and\ \bibinfo {author} {\bibfnamefont {M.~R.}\ \bibnamefont {Visser}},\ }\bibfield  {title} {\bibinfo {title} {{Holographic dictionary for Lifshitz and hyperscaling violating black holes}},\ }\href@noop {} {\  (\bibinfo {year} {2024})},\ \Eprint {https://arxiv.org/abs/2410.16145} {arXiv:2410.16145 [hep-th]} \BibitemShut {NoStop}%
\bibitem [{\citenamefont {Kiritsis}\ and\ \citenamefont {Matsuo}(2017)}]{Kiritsis:2016rcb}%
  \BibitemOpen
  \bibfield  {author} {\bibinfo {author} {\bibfnamefont {E.}~\bibnamefont {Kiritsis}}\ and\ \bibinfo {author} {\bibfnamefont {Y.}~\bibnamefont {Matsuo}},\ }\bibfield  {title} {\bibinfo {title} {{Hyperscaling-Violating Lifshitz hydrodynamics from black-holes: Part II}},\ }\href {https://doi.org/10.1007/JHEP03(2017)041} {\bibfield  {journal} {\bibinfo  {journal} {JHEP}\ }\textbf {\bibinfo {volume} {03}},\ \bibinfo {pages} {041}},\ \Eprint {https://arxiv.org/abs/1611.04773} {arXiv:1611.04773 [hep-th]} \BibitemShut {NoStop}%
\bibitem [{\citenamefont {Kim}\ \emph {et~al.}(2013)\citenamefont {Kim}, \citenamefont {Kulkarni},\ and\ \citenamefont {Yi}}]{Kim:2013zha}%
  \BibitemOpen
  \bibfield  {author} {\bibinfo {author} {\bibfnamefont {W.}~\bibnamefont {Kim}}, \bibinfo {author} {\bibfnamefont {S.}~\bibnamefont {Kulkarni}},\ and\ \bibinfo {author} {\bibfnamefont {S.-H.}\ \bibnamefont {Yi}},\ }\bibfield  {title} {\bibinfo {title} {{Quasilocal Conserved Charges in a Covariant Theory of Gravity}},\ }\href {https://doi.org/10.1103/PhysRevLett.111.081101} {\bibfield  {journal} {\bibinfo  {journal} {Phys. Rev. Lett.}\ }\textbf {\bibinfo {volume} {111}},\ \bibinfo {pages} {081101} (\bibinfo {year} {2013})},\ \bibinfo {note} {[Erratum: Phys.Rev.Lett. 112, 079902 (2014)]},\ \Eprint {https://arxiv.org/abs/1306.2138} {arXiv:1306.2138 [hep-th]} \BibitemShut {NoStop}%
\bibitem [{\citenamefont {Gim}\ \emph {et~al.}(2014)\citenamefont {Gim}, \citenamefont {Kim},\ and\ \citenamefont {Yi}}]{Gim:2014nba}%
  \BibitemOpen
  \bibfield  {author} {\bibinfo {author} {\bibfnamefont {Y.}~\bibnamefont {Gim}}, \bibinfo {author} {\bibfnamefont {W.}~\bibnamefont {Kim}},\ and\ \bibinfo {author} {\bibfnamefont {S.-H.}\ \bibnamefont {Yi}},\ }\bibfield  {title} {\bibinfo {title} {{The first law of thermodynamics in Lifshitz black holes revisited}},\ }\href {https://doi.org/10.1007/JHEP07(2014)002} {\bibfield  {journal} {\bibinfo  {journal} {JHEP}\ }\textbf {\bibinfo {volume} {07}},\ \bibinfo {pages} {002}},\ \Eprint {https://arxiv.org/abs/1403.4704} {arXiv:1403.4704 [hep-th]} \BibitemShut {NoStop}%
\end{thebibliography}

%

\end{document}